\documentclass[fleqn,usenatbib]{mnras}

\usepackage{mathptmx}

\usepackage[T1]{fontenc}
\usepackage{ae,aecompl}
\usepackage{color}

\usepackage{graphicx}	
\usepackage{amsmath}	
\usepackage{amssymb}	

\title[Tidal Disruption of Planets and BDs in AGB Stars]{The Formation of Discs in the Interior of AGB Stars from the Tidal Disruption of Planets and Brown Dwarfs}

\author[G. Guidarelli et al.]
{G. Guidarelli$^{1}$\thanks{gcg3642@g.rit.edu},
J. Nordhaus$^{1,2}$\thanks{nordhaus@astro.rit.edu},
J. Carroll-Nellenback$^{3}$,
L. Chamanady$^{3}$,
A. Frank$^{3}$,
\newauthor E. G. Blackman$^{3}$
\\
$^{1}$Center for Computational Relativity and Gravitation, Rochester Institute of Technology, Rochester, NY 14623, USA\\
$^{2}$National Technical Institute for the Deaf, Rochester Institute of Technology, Rochester, NY 14623, USA\\
$^{3}$Department of Physics and Astronomy, University of Rochester, Rochester, NY 14627, USA
}
\date{Accepted XXX. Received YYY; in original form ZZZ}

\pubyear{2021}

\begin{document}
\label{firstpage}
\pagerange{\pageref{firstpage}--\pageref{lastpage}}
\maketitle

\begin{abstract}
A significant fraction of isolated white dwarfs host magnetic fields in excess of a MegaGauss.  Observations suggest that these fields originate in interacting binary systems where the companion is destroyed thus leaving a singular, highly-magnetized white dwarf. In post-main-sequence evolution, radial expansion of the parent star may cause orbiting companions to become engulfed.  During the common envelope phase, as the orbital separation rapidly decreases, low-mass companions will tidally disrupt as they approach the giant's core.  We hydrodynamically simulate the tidal disruption of planets and brown dwarfs, and the subsequent accretion disc formation, in the interior of an asymptotic giant branch star. These dynamically formed discs are commensurate with previous estimates, suggesting strong magnetic fields may originate from these tidal disruption events.
\end{abstract}



\section{Introduction}

White dwarfs (WDs) with magnetic field strengths in excess of a MegaGauss (MG) are observed in both isolated and binary systems \citep{Ferrario2015}. WDs in binaries can either be "attached" (the companion's radius extends past the equipotential surface and mass transfer is occurring through L1) or "detached" (the companion's radius is well within the equipotential surface). Between 25\% and 36\%  of WDs in attached binary systems are highly magnetic (polar or intermediate polar cataclysmic variables) while the remaining systems have sub-MG or non-measurable fields \citep{Ferrario2015, Pala_2020}. In contrast, WDs in detached binary systems seldom host strong magnetic fields with only $\sim$10 of the over 3000 found to date \citep{Rebassa_2016, Parsons_2021}. Lastly, about 10\% of isolated WDs have measured field strengths in excess of a MG, these objects are commonly referred to as High Field Magnetic White Dwarfs (HFMWDs) \citep{Kawka_2007,Liebert_2003}. In each of these cases the origin of the WD's magnetic field strength is an area of active research.

 HFMWDs tend to be more massive than their isolated non-magnetic counterparts \citep{Liebert_1988}. This suggests that HFMWD formation may require some type of mass transfer during binary interactions. Additionally, a primordial field origin in which magnetized white dwarfs are descendants of Ap/Bp stars is unable to explain the quantity and field strength distribution of HFMWDs in the solar neighborhood \citep{Kawka_2004, kawka_2020}.  If the origin were truly independent of binary interactions, then the presence of a detached binary should not affect the incidence of magnetic fields. However, in contrast to the $\sim$$10\%$ of isolated white dwarfs that host strong magnetic fields, not a single young magnetic white dwarf has been observed in a detached binary\footnote{The systems in \cite{Rebassa_2016} and \cite{Parsons_2021} all contain older WDs and companions with high Roche-lobe filling factors. The latter may indicate a history of mass transfer, i.e., not necessarily detached.} \citep{Kawka_2007,Liebert_2003}. Therefore, the formation channel for strong magnetic fields in WDs likely requires binary evolution \citep{Tout_2008,Nordhaus:2011aa}.

Recently, research on binary formation  for magnetic white dwarfs has focused on pathways that  involve common envelope evolution (CEE) \citep{Tout_2008,Nordhaus:2011aa,FERRARIO20201025}. In CEE,
one of the stellar components experiences radial expansion, creating an envelope of loosely bound material.  This envelope may directly engulf an orbiting companion or cause orbital decay that results in a common envelope (CE) phase \citep{Paczynski1976,Harpaz_1994,Siess_1999,Nordhaus2006,Staff_2016,Belczynski:2016sf,wilson2019,2020MNRAS.497.1895W}. For a more complete discussion of CEs see \cite{Ivanova:2013aa} and references therein. Magnetic field generation could occur in the envelope during CEE, or during subsequent mass transfer event in the post-CE system, or in a merger if the companion disrupts during the CE phase \citep{Tout_2008,Nordhaus:2007aa,Nordhaus:2011aa,Ohlmann_2016}.

Double degenerate (WD-WD) mergers are one proposed pathway to create isolated HFMWDs \citep{Wickramasinghe_2000,Garc_a_Berro_2012}. In this scenario, the resultant object is a massive white dwarf with a hot differentially rotating corona that may amplify magnetic fields via a dynamo.  This seems unlikely to explain the complete distribution of isolated HFMWDs, as the average mass of HFMWDs is only about 0.1 $\mathrm{M_\odot}$ more than the average mass of non-magnetic WDs. Furthermore, a small number of HFMWDs have been discovered in young stellar clusters.  In these settings, only the most massive WDs would exist (The remnants of the most massive intermediate mass stars), and therefore mergers of two of these objects would have resulted in Type Ia supernovae \citep{Caiazzo_2020}.

Here, we focus on CE phases that result in the merger of a white dwarf with a less massive companion (M-dwarf, brown dwarf, or planetary companion) \citep{Nordhaus:2011aa,Nordhaus:2013aa,Guidarelli2019}. During CEE, the companion does not possess sufficient energy to eject the envelope and emerge in a short-period orbit.  Instead, the orbital separation decreases until the companion is tidally disrupted as it nears the proto-white dwarf core.  The disrupted material forms an accretion disc that may amplify the magnetic field and advect it to the white dwarf surface where it could anchor.

Previously, we showed that such discs survive thermal evaporation in the hot stellar ambient medium and remain stable for at least $\sim$$100$ orbits \citep{Guidarelli2019}. However, these results assumed a disc formed from disrupted material was already present and thus are contingent on assumptions about the initial mass and disc geometry we chose. To verify that these assumptions are reasonable and to obtain further insight, we numerically simulate the dynamics of the tidal disruption events and subsequent disc formation. Some work has been done on massive star binary mergers with low-mass companions, investigating the long term evolution of angular momentum as well as the surface abundances \citep{Manos_2020,Wu_2020}. Tidal disruption events around white dwarfs have also been investigated \citep{Malamud_2020,Veras_2014}. However, there have been no simulations investigating disc formation and structure from tidal disruption around the core of an Asymptotic Giant Branch (AGB) star. 

In this paper we present high-resolution, three-dimensional, adaptive mesh refinement (AMR) simulations of tidal disruption events involving $10$, $20$, and $30$ ${\rm M_J}$ \footnote{$\mathrm{M_{\rm J}}$ is the mass of Jupiter} companions inside the core of an AGB star that was evolved from a 2 $\mathrm{M_\odot}$ zero-age-main-sequence star.  We find that an accretion disc typically forms within six orbital periods at $\sim$$\mathrm{1 R_\odot}$.  Approximately, $\sim$60\% of the disrupted material orbits in a disk
while the rest remains gravitationally bound in the tidal tail.
Even though the disc dynamically forms, its resultant structure is similar to the initial conditions assumed in our previous work \citep{Guidarelli2019}.

The structure of the paper is as follows: in Section 2, we describe our numerical techniques and initial conditions. In Section 3, we analyze the evolution of the global angular momentum, present the resultant radially averaged angular velocity profile of the disc, quantify the fraction of tidally-disrupted mass that forms the disc, analyze the flow properties and discuss how our results depend on resolution. We discuss future work and conclude in Section 4.

\section{Numerical Setup}

We perform 3D hydrodynamic simulations of Tidal Disruption Events (TDEs) with $\mathrm{10\ ,\ 20, \ and \ 30} {\rm \ M_{\rm J}}$ companions around the ${\rm 0.56\ M_\odot}$ core of an AGB star. In contrast to CE simulations that utilize the entire star \citep{Sands_2020,Chamandy_2019,Chamandy_2019b, Chamandy_2020,Ohlmann_2015,Ohlmann2017,DeMarco2011, Ricker_2008, Ricker2012}, we focus on the inner $10^{12}$ cm to study the dynamics of the tidal disruption event . This results in a higher-effective resolution than what is typically achieved in global CE simulations that simulate the entire primary star.  

The companions were initialized just outside the tidal disruption radius. We define this as the distance of the secondary from the center of mass of the primary (AGB core) where the companion's gravitational binding energy is equal to the average differential external potential multiplied by the diameter of the planet. This separation is given as: 

\begin{equation}
    \mathrm{R_{tidal} \sim \left(\frac{2(5-n)}{3} \frac{M_{core}}{M_p}\right)^{1/2} R_p}\ ,
\end{equation}

where $\mathrm{R_p}$ and $\mathrm{M_p}$ are the companion radius and mass, and $\mathrm{M_{core}}$ is the mass of the proto-WD AGB core.  This assumes that the planet is a polytrope with index $n$, and the companion's radius is small compared to the separation.   

The companions were given an initial orbital velocity that was 65\% of the Keplerian velocity at that radius, this conserves computational resources as it ensures tidal disruption within one orbit. The simulations were run for over 100 orbits of the inner-most regions of the disc after it formed, and at least six orbits of the disc at $\sim$$\mathrm{1 R_\odot}$ (i.e. the approximate outer edge of the disc).  We investigate the dynamics of the tidally disrupted matter, the fraction of initial mass that forms the disc, the disc geometry, and the timescale of disc formation. Details of the numerical grid, initial conditions, modifications to the stellar profile, and other aspects of the numerical implementation are described below.

\subsection{Numerical Grid Parameters}

To carry out these simulations we utilize AstroBEAR, an Eulerian, three-dimensional (3D) multi-physics code \citep{Cunningham_2009,CARROLLNELLENBACK_2013}. AstroBEAR employs adaptive mesh refinement (AMR) to dynamically resolve regions of the grid that require higher resolution while leaving other regions at lower resolution, thereby saving substantial computational costs.

AstroBEAR solves the following hydrodynamic equations: 

\begin{equation}
    \mathrm{\frac{\partial \rho}{\partial t} + \nabla \cdot (\rho \mathbf{v}) = 0},
\end{equation}
\begin{equation}
     \frac{\partial \rho \mathbf{v}}{\partial t} + \nabla \cdot (\rho \mathbf{v v}) = -\nabla P - \rho g_c(r) \hat{\textbf{r}} + \rho \textbf{a}_g,
\end{equation}
\begin{equation}
    \frac{\partial E}{\partial t} + \nabla \cdot [(E + P)\mathbf{v}] = \rho(-g_c(r) \hat{\textbf{r}} +  \textbf{a}_g)\cdot \mathbf{v},
\end{equation}

\begin{equation}
    E = \frac{1}{2}\rho(\mathbf{v\cdot v}) + \mathcal{E}
\end{equation} 

and
\begin{equation}
     P = \tilde{n} k_\mathrm{b} T,
\end{equation}
where $\rho$ is the density, $t$ is time, $\mathbf{v}$ is the velocity, $P$ is the pressure, $r$ is the radial distance to a point particle, $\textbf{a}_g$ is the acceleration due to self gravity, $\mathcal{E}$ is the total internal energy density, $\tilde{n}$ is the number density of particles, $k_\mathrm{b}$ is the Boltzmann constant and $T$ is the temperature. For these simulations we use an ideal gas equation of state with adiabatic index, $\gamma = 5/3$. Lastly, $g_c$ is the softened gravitational acceleration from a point particle :

\begin{align} \label{gc}
g_c(r) = 	\mathrm{G M_{pp}} \left\{ \begin{array}{cc} 
                r^{-2} & \hspace{3mm} r\ge h \\
                \\
                \frac{y\left(\frac{64}{3}+y\left(-48+y\left(\frac{192}{5}+y\frac{-32}{3}\right)\right)\right)-\frac{1}{15y^2}}{h^{2}} & \hspace{3mm} h/2\le r<h \\
                \\
               \frac{y\left(\frac{32}{3}+y^2\left(\frac{-192}{5}+32y\right)\right)}{h^2} & \hspace{3mm} r<h/2 \\
                \end{array} \right.
\end{align}
where h is the softening radius, and $y = r/h$. $\mathrm{M_{pp}}$ is the mass of the point particle which is smaller than $\mathrm{M_{core}}$. The difference in mass $\mathrm{M_{core} - M_{pp}}$ is taken into account with $\textbf{a}_g$ as it is resolved on the grid (see Section 2.2.1).

We employ a base grid of $32^3$ with up to five levels of factor two refinement for an effective resolution of $1024^3$. This yields a smallest grid cell of $\sim$$10^9$ cm, approximately the radius of the central white dwarf. The total simulated box size is $(2\times10^{12} \mathrm{cm})^3$. The spline softening radius of the central particle is $10^{10}$ cm. We simulate approximately one day of physical time, consistent with six orbits at the outer radii of the resultant disc and over 100 orbits at the softening radius.

\begin{figure}
    \includegraphics[scale = 0.55]{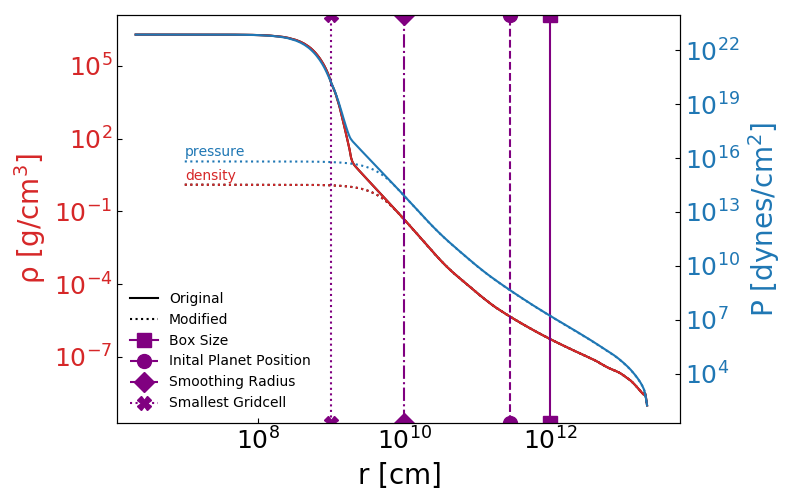} 
    \caption{AGB density and pressure profiles before (solid line) and after (dotted line) modification. The vertical lines highlight the simulation box size, the initial companion separation, the gravitational smoothing radius, and the smallest grid cell. Note that after modification, the smallest grid cell aligns with a much smaller density and pressure gradients.}
    \label{fig:profiles}
\end{figure}

\begin{figure}
    \centering
    \includegraphics[scale = 0.28]{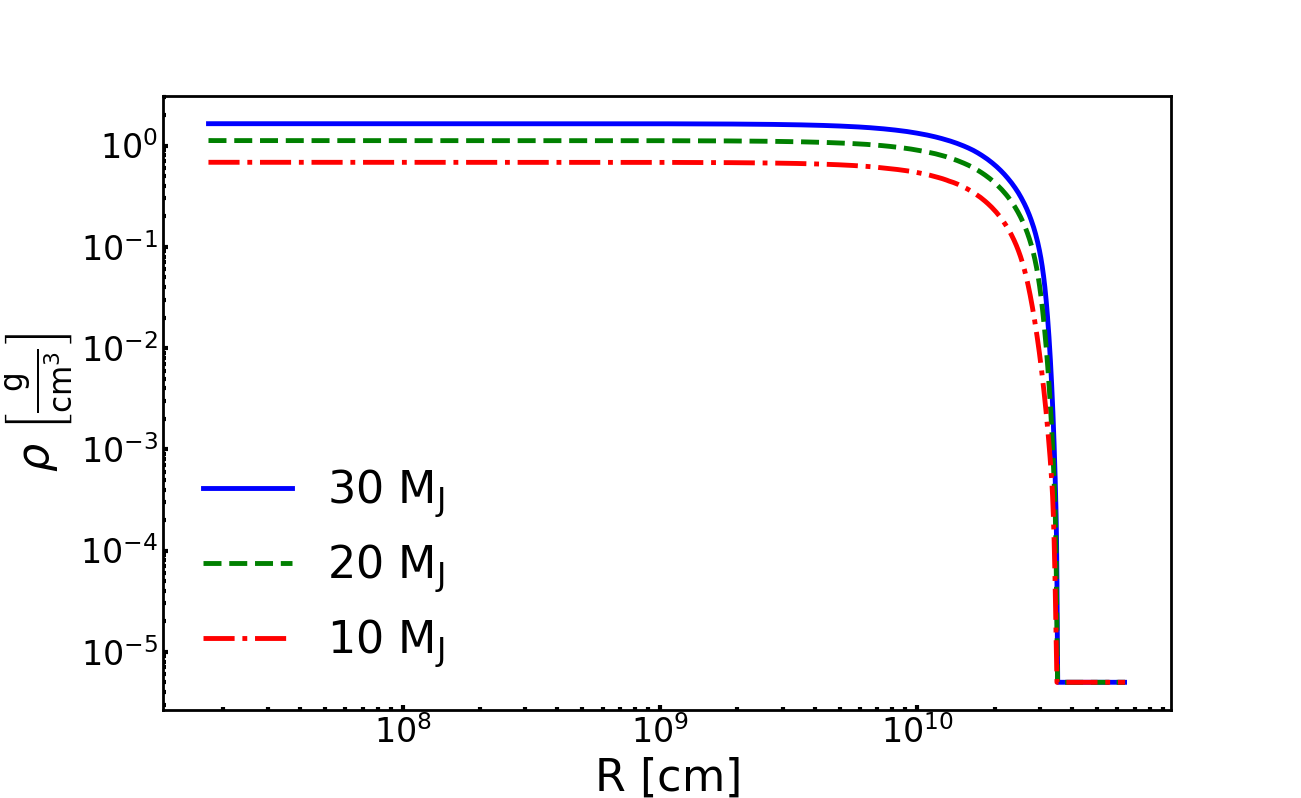}
    \caption{Density as a function of radius for the planetary companions, having an effective polytropic index, $n \sim 1.5$. Because the profiles are already flat within $10^9$ cm (the smallest grid cell), they do not require a point particle or modification. The density corresponding to the flat outer section is the average AGB density in the companion vicinity. This section is appended to the profile to create an initial co-moving envelope of relatively low density material which reduces the effects of the shock between the orbiting companion and stationary ambient material.}
    \label{fig:planetProfiles}
\end{figure}

\begin{figure*}
    \centering
    \includegraphics[height=6cm]{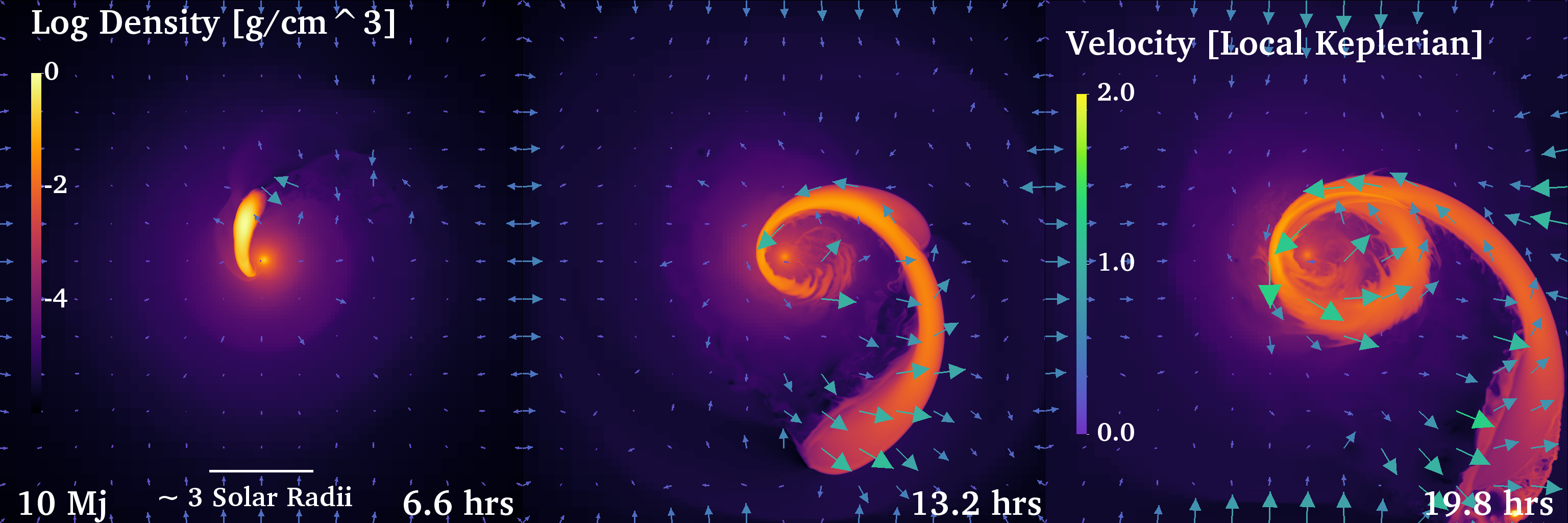}
    \caption{Evolution of the 10 Jupiter mass disc at $6.6$, $13.2$ and $\mathrm{19.8\ hrs}$. Each panel shows the density distribution in the orbital plane indicated by colour. Velocity vectors are shown on all panels with magnitude indicated by colour (in units of the local Keplerian velocity) and the projected magnitude by the length.}
    \label{fig:10MJ}
\end{figure*}

\begin{figure*}
    \centering
    \includegraphics[height=6cm]{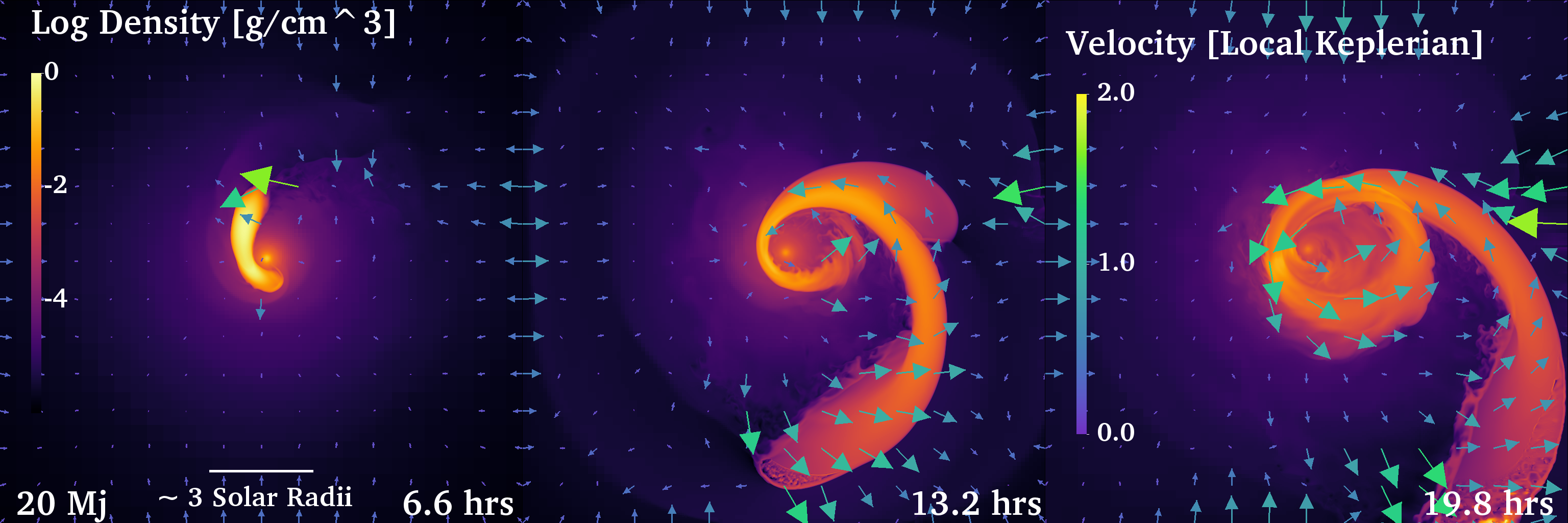}
    \caption{Evolution of the 20 Jupiter mass disc at $6.6$, $13.2$ and $\mathrm{19.8\ hrs}$. Similar to Figure~\ref{fig:10MJ}}
    \label{fig:20MJ}
\end{figure*}

\begin{figure*}
    \centering
    \includegraphics[height=6cm]{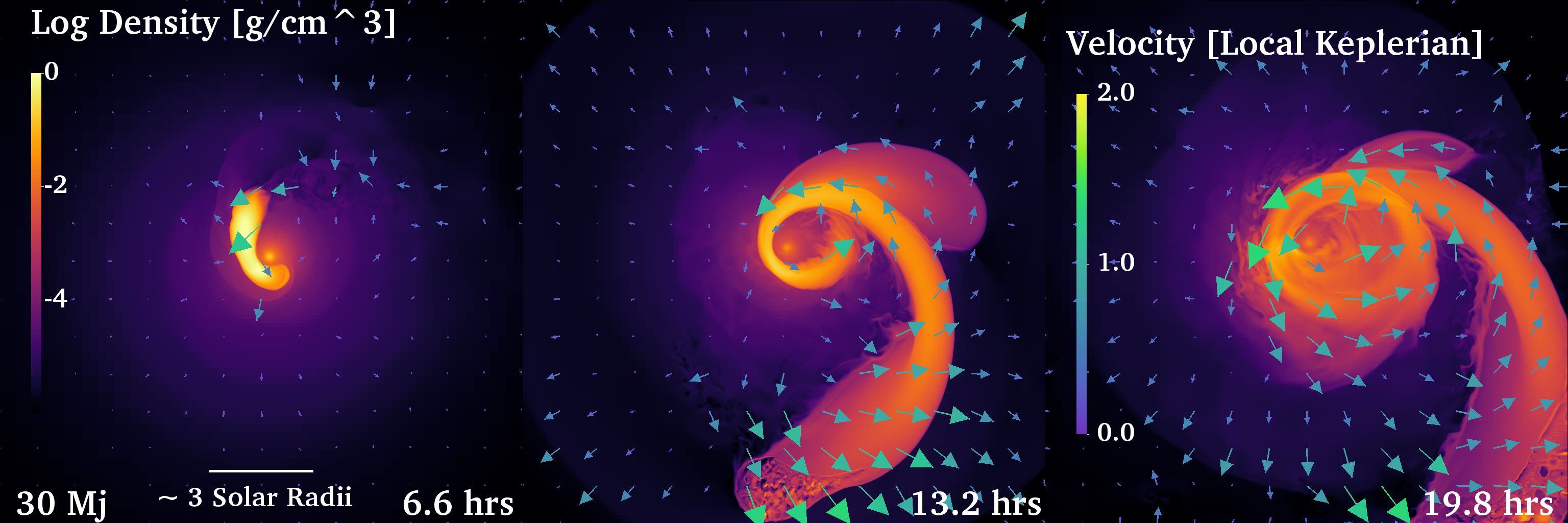}
    \caption{Evolution of the 30 Jupiter mass disc at $6.6$, $13.2$ and $\mathrm{19.8\ hrs}$. Similar to Figure~\ref{fig:10MJ}}
    \label{fig:30MJ}
\end{figure*}

\begin{figure}
    \centering
    \includegraphics[scale = 0.25]{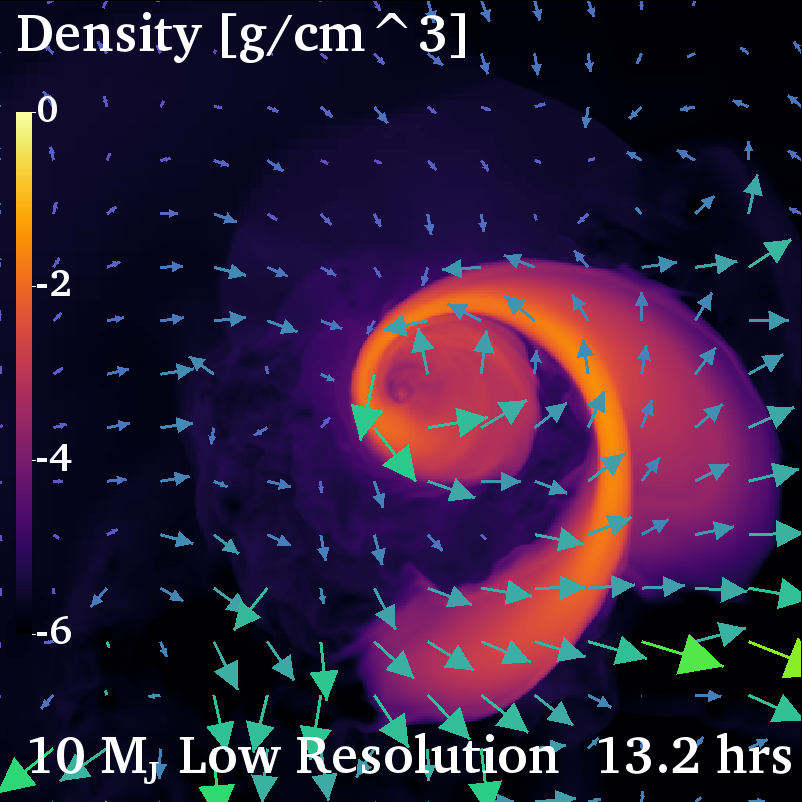}
    \caption{Lower resolution version of our $\mathrm{10 M_J}$ simulation. Although more dispersive, the results are morphological similar  over the simulated time. The arrows represent velocity and use the same color scale as in Figures \ref{fig:10MJ}-\ref{fig:30MJ}.}
    \label{fig:lowres}
\end{figure}

\begin{figure}
    \centering
    \includegraphics[scale = 0.45]{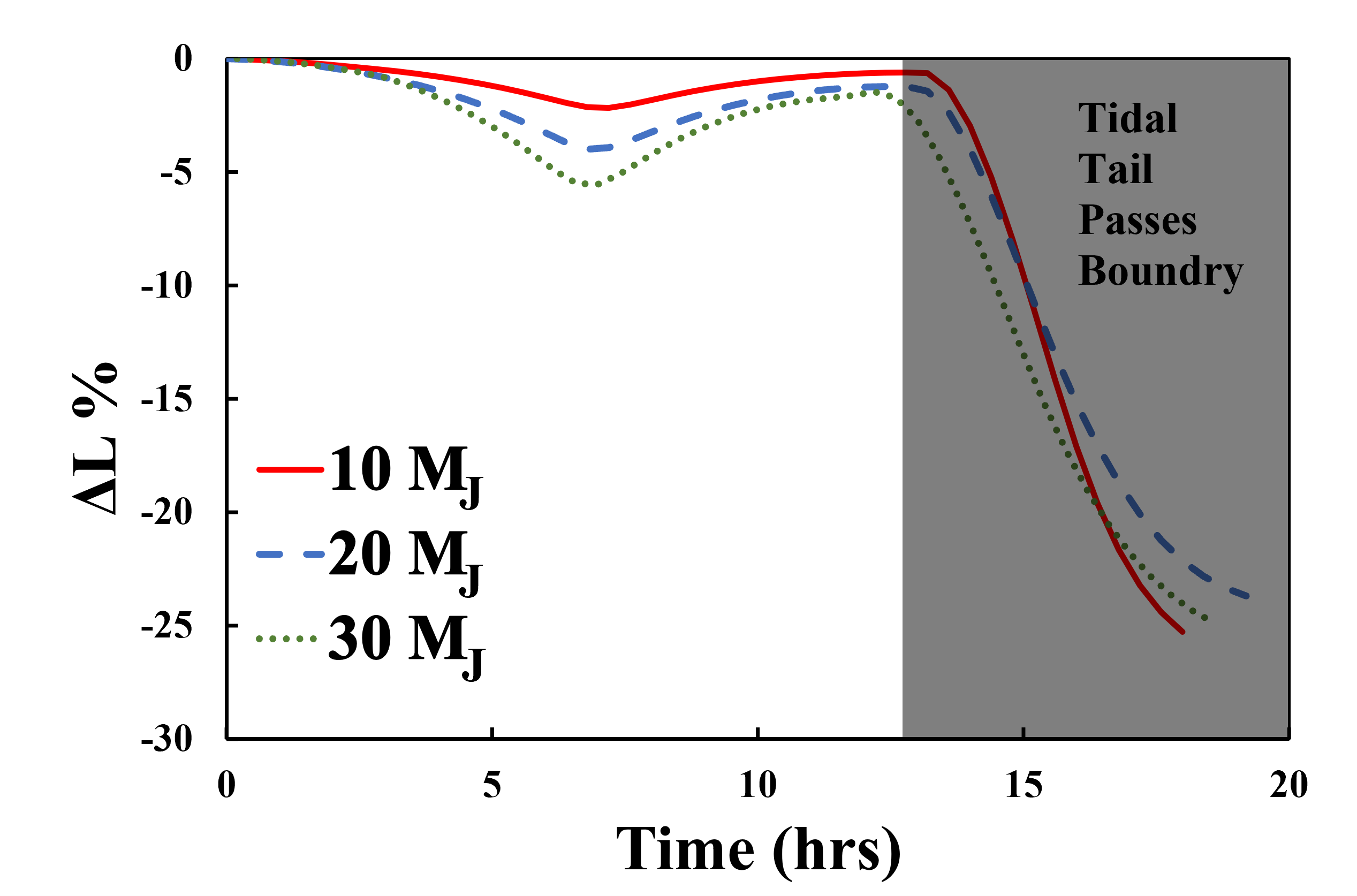}
    \caption{The total angular momentum as a function of time within a cylinder of diameter $80\%$ the simulation box side length. The $\sim$$5\%$ dip in angular momentum at $\sim$$6\ \mathrm{hr}$ is caused by a small amount of material accelerated by the companion that exits the cylinder only to slow down and fall back in the the following few hours.  As indicated by the gray shaded region, At $13\ \mathrm{hr}$, the tidal arm passes through the cylinder boundary and causes the total angular momentum within the cylinder to decrease accordingly.}
    \label{fig:angular}
\end{figure}

\begin{figure}
    \centering
    \includegraphics[scale = 0.35]{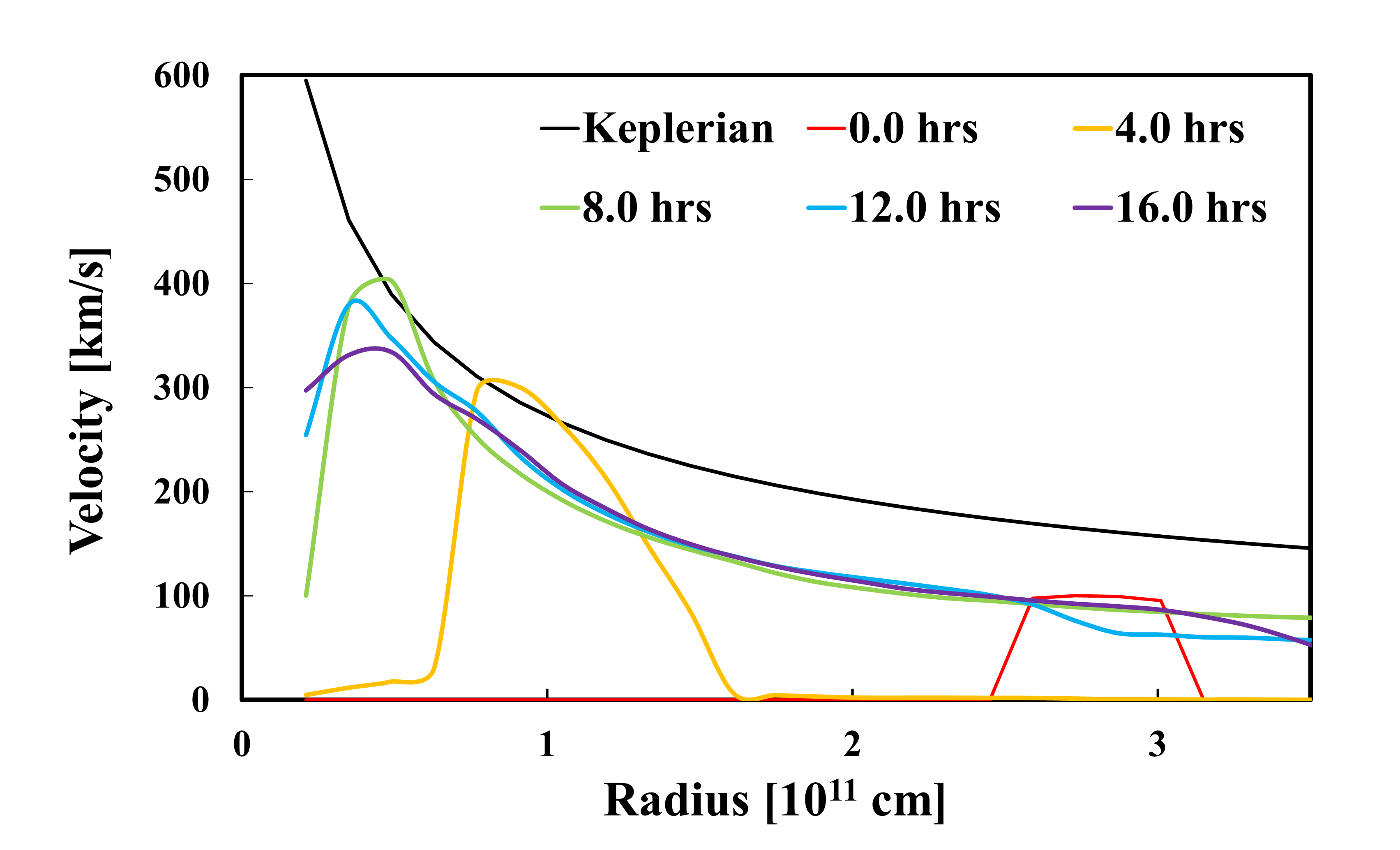}
    \caption{Azimuthally averaged momentum per unit mass perpendicular to radial position at different times of the $10 \mathrm{M_J}$ simulation. The black curve is the Keplerian velocity profile about a point particle of mass $\mathrm{0.565\ M_\odot\ (M_{pp})}$. While our simulations result in elliptical discs, we expect circularization in a few more orbital time scales. When this occurs, the disc will still be sub-Keplerian as pressure gradients will offset some gravitational force.}
    \label{fig:vel}
\end{figure}

\begin{figure*}
    \centering
    \includegraphics[scale = 0.27]{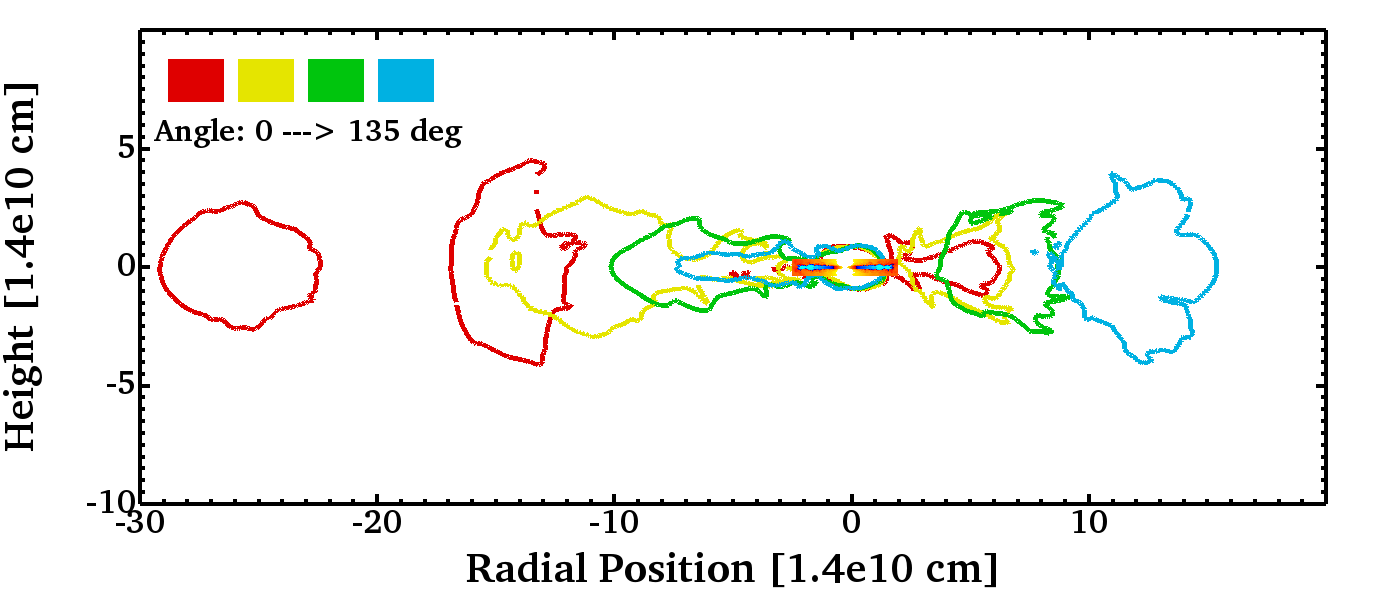}
    \caption{Contours of constant density ($\mathrm{0.02\ \mathrm{g/cm^3}}$) in planes perpendicular to the orbital plane at $\mathrm{20\ hrs}$ into the $10\  \mathrm{M_J}$ simulation. Color indicates the azimuthal angle in increments of 45 degrees about the point particle. Superimposed is a to-scale image of the $10\ \mathrm{M_J}$ disc from our previous work \citep{Guidarelli2019}. The red blob on the left is a cross-section of the tidal tail.}
    \label{fig:contours}
\end{figure*}

\begin{figure}
    \centering
    \includegraphics[scale = 0.15]{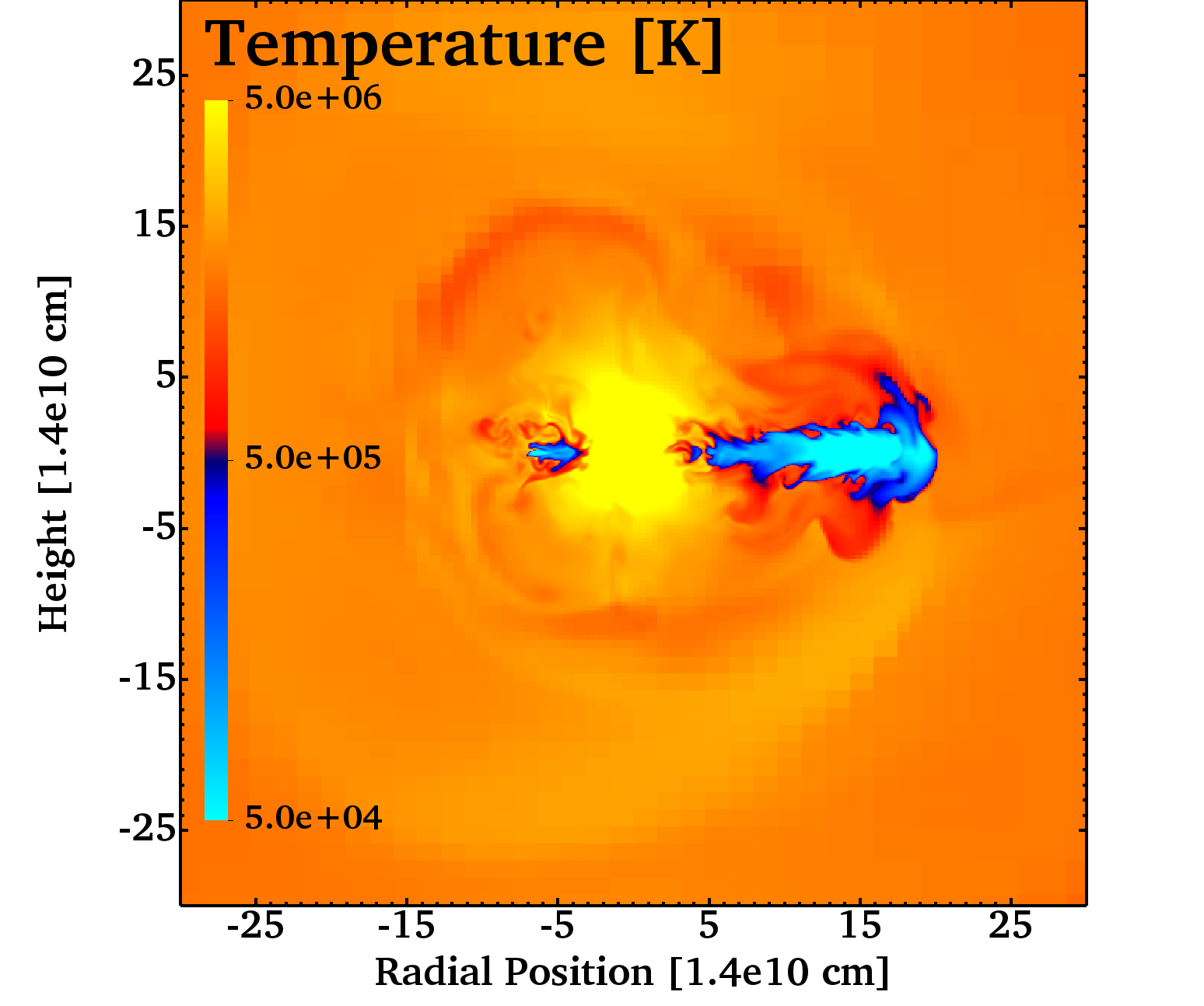}
    \caption{Vertical slice through the orbital plane showing the internal temperature of the disc at 26.4 hrs into the $\mathrm{10\ M_J}$ companion simulation. The disc is much colder than the stellar interior and entrainment of the of ambient material is caused by the intense vertical shear.}
    \label{fig:temp}
\end{figure}

\subsection{Initial Condition Generation and Modification}

To generate interior profiles for both the AGB star and the companions, we utilize MESA (Modules for Experiments in Stellar Astrophysics version 12115) \citep{Paxton:2011aa,Paxton:2018aa,Paxton_2019}. MESA is a spherically symmetric, stellar evolution code that is capable of simulating the interior density, pressure, temperature and composition profiles of a star over various epochs of its evolution.  We calculate the full evolution of a primary star with a zero-age-main-sequence mass of 2 ${\rm M_\odot}$ with solar metallicity.  Mass-loss on the Red Giant Branch (RGB) is governed via a Reimers mass-loss prescription with a Reimers mass-loss coefficient, $\eta_{\rm R}=0.7$ \citep{1975MSRSL...8..369R}.  On the AGB, we use a Bloecker mass-loss prescription with a mass-loss coefficient, $\eta_{\rm B}$, of 0.7 \citep{1995A&A...297..727B}.  For the simulations reported in this paper, we select the density and pressure profiles when the radius attains its maximum extent on the AGB. These spherically-symmetric profiles are mapped onto the grid and form the initial conditions for the AGB ambient. 

Similarly, we use {\sc MESA} to generate hydrostatic density and pressure profiles for each planet and brown dwarf companion. With the built in \textit{'make\_planets'} module, one can create an initial 1D hydrostatic profile, allow this profile to relax and form a core, and expose the profile to expected radiation. We initialize our companion with solar metallicity values of $Y = 0.25$ and $Z = 0.02$ and select these profiles before cooling and core formation which results in a slightly larger estimate for the tidal disruption radius. Note that a typical core contributes $<10\%$ of the planet mass.

\subsubsection{AGB Profile Modification}

Because of the large range of scales inherent to AGB stars, we utilize a point particle to represent some of the mass in the central region. Were a point particle  not employed, the steep density gradients in this region would be under-resolved, resulting in numerical instability on small scales and an inaccurate gravitational potential at larger scales. Once implemented, point particles allow for the original steep density profile to be replaced with a flatter profile. However, point particles also require a modification to the gravitational potential as under-resolved small separations lead to extraneous acceleration. Therefore, to maintain a proper representation of the potential when implementing a point particle, we modify the density 
profile in accordance with a smoothed gravitational potential  acting on the surrounding material. This modified profile is shown in Figure \ref{fig:profiles}. 

We modify the density and pressure distributions using a method previously described in \cite{Guidarelli2019}. Specifically, we construct a numerical integrator to iterate over solutions to the Modified-Lane-Emden equation. This yields a modified density and corresponding pressure profile that maintains hydrostatic equilibrium for any choice of initial profile and smoothing function.  This code is publicly available\footnote{\url{https://github.com/GuidarelliG/MLERK4}} and is useful for generating hydrostatic initial conditions for 3D simulations involving evolved stars \citep{Guidarelli2019, Ohlmann2017,Chamandy:2018aa}.

\subsubsection{Planet and brown dwarf co-moving envelopes}

In contrast to the AGB initial conditions, the planetary profiles could be imported to the 3D grid without modification, as the smallest grid-cell accurately resolves all density gradients. However, because the pressure gradients at the planetary boundary cause large initial shocks, we give the surrounding ambient (a spherical shell of thickness $\mathrm{0.5}\ R_p$) the same initial velocity as shown in Figure \ref{fig:planetProfiles}. While shocks are expected, the initial steep pressure gradient significantly lengths the computational time.  By adding this low-density shell of co-moving material, the initial shock dissipates quickly allowing subsequent shocks to develop naturally. 

\section{Results}

Figures \ref{fig:10MJ} - \ref{fig:30MJ} present the density distributions in the orbital plane from each simulation at $6.6$, $13.2$ and $\mathrm{19.8\ hrs}$. These times correspond to $\sim$$2$, 4 and 6 Keplerian orbits at the outer disc radii. In each panel we also show the velocity in units of the local Keplerian circular speed. From these images, it is clear a disc begins to form within $\mathrm{20\ hrs}$ regardless of the initial companion mass. Within the first few hours the planets disrupt, dispersing the angular momentum. The material that gains angular momentum forms a tidal tail and moves outward while material that loses significant angular momentum falls inward. The main differences between these simulations stem from the increase in total linear momentum. During disruption, the more massive companions transfer more linear momentum to the primary core. Because the core then moves in the direction of this momentum, material in the opposite direction becomes more dispersed.

Thermal evaporation seems to play a minor role as the disrupted companion mass remains relatively distinct from the hot stellar interior.  At the end of the simulations, $\sim$$60\%$ of the initial companion mass resides in the disc.  The material in the tidal tail remains bound and may fall back on longer timescales.

To quantify the effect of resolution on our results, we present a snapshot from an identical $10\  \mathrm{M_J}$ simulation at lower resolution in Figure \ref{fig:lowres}. Although more dispersive, the lower resolution simulation is dynamically and morphologically consistent with the higher resolution run.

\subsection{Angular Momentum Evolution}

The total angular momentum within a cylinder of radius $\mathrm{0.8 \times10^{12} cm}$ is shown in Figure \ref{fig:angular} as a function of time for each of the simulations. In all cases angular momentum is conserved to within $\sim$$5\%$ until material from the disrupted companion leaves the computational grid at $\sim$$\mathrm{13 hrs}$.

The distribution of angular velocity for the 10 $M_{\rm J}$ planet is shown in Figure \ref{fig:vel}. Each color represents a different time in the simulation in reference to the Keplerian velocity (black line). As the planet moves inwards, density and angular momentum diffuse radially. By the end of the simulation, the angular velocity profile matches the concavity of the Keplerian solution. The difference in magnitude is the result of a non-circular (elliptical) disc as well as pressure support within the disc.

\subsection{Resultant Disc Geometry and Efficiency}

Figure \ref{fig:contours} presents density contours of the disc taken from slices through the point particle perpendicular to the orbital plane at various angles (indicated by color). In this figure, we included an image of a $10\mathrm{M_J}$ disc from our previous work for comparison \citep{Guidarelli2019}. In aggregate, the contours provide an estimate of the height to radius ratio of the disc. We find that for these discs $H/R$ varies from $0.05$ in the inner regions to $0.25$ in the tidal tail consistent with our previous work \citep{Guidarelli2019}. The temperature in a vertical slice through the disc is shown in Figure \ref{fig:temp} and demonstrates that post-tidal disruption, the disk remains distinct and cold compared to the hot ambient.

\section{Conclusions}

In this work, we present the results of 3D hydrodynamic simulations of tidal disruption events of planet and brown dwarf companions in the interior of an AGB star. These events are expected to form accretion discs which may result in conditions that amplify and anchor strong magnetic fields to the proto-WD \citep{Nordhaus:2010aa}. As mass-loss from the AGB star continues during its evolution, a HFMWD would eventually emerge. We show that for planetary and brown dwarf companions, tidal disruption results in the formation of an accretion disc around the proto-WD which typically occurs within a few orbital timescales. About 60\% of the original companion mass forms the disc with the remaining 40\% in the tidal tail and subject to fallback on longer timescales. Lastly, the innermost region of our tidal discs are morphologically similar to that of the disc initial conditions assumed in our previous work \citep{Guidarelli2019}. 

Once the disc is formed, advection of magnetic flux through the disc or its surface might transport the field to the white dwarf. Note that this would occur
on radial accretion timescales\footnote{The radial accretion time 
for a standard disk can be estimated as:  
$\sim$$(R/ H)^2 (1/\alpha\Omega)$ where R is the disc radius, H is the disc height, $\Omega$ is the rotation frequency, and $\alpha$ is the viscous coefficient associated with Shakura and Sunyaev type discs.} ($\leq$ 1 month) for a standard disk, which is much shorter then a typical AGB lifetime ($\sim$$10^{5-6}$ years). 

Within the environment of a stellar envelope however, a buildup of pressure may halt accretion in the absence of a pressure release valve. A pressure release valve could be provided by jets \citep{Chamandy:2018aa}. The magnetic field itself  may incite vertical outflows which act to relieve pressure, in essence forming a self-sustaining conveyor belt of magnetic flux until the field gets large enough to inhibit further advection. Future work should therefore include a study of the dynamics of tidal disruption when the companion, and/or stellar ambient, are initially magnetized. 

\section*{Acknowledgements}

GG and JN acknowledges support from the following grants: NSF AST-2009713, NASA HST-AR-15044 and NTID SPDI-15992.  This work used the Extreme Science and Engineering Discovery Environment (XSEDE), which is supported by National Science Foundation grant number ACI-1548562. Financial support for AstroBEAR is currently provided by the Department of Energy grants DE-SC0001063, DE-SC0020432 and, DE-SC0020434 the National Science Foundation grant AST-1813298, and the Space Telescope Science Institute grant HST-AR-12832.01-A. The Center for Integrated Research Computing (CIRC) at the University of Rochester provided additional computational and visualization resources. {\rm VisIt} is supported by the Department of Energy with funding from the Advanced Simulation and Computing Program and the Scientific Discovery through Advanced Computing Program. 





\bibliographystyle{mnras}
\bibliography{MNRASpaper} 




\bsp	
\label{lastpage}
\end{document}